\documentclass[aps,prl,twocolumn,superscriptaddress,nofootinbib,longbibliography]{revtex4-2}

\usepackage[T1]{fontenc}
\usepackage[utf8]{inputenc}
\usepackage{lmodern}
\usepackage{microtype}
\usepackage{amsmath,amssymb,mathtools,amsthm}
\usepackage{bm}
\usepackage{braket}
\usepackage{hyperref}
\usepackage[dvipsnames]{xcolor}
\hypersetup{colorlinks=true,linkcolor=MidnightBlue,citecolor=MidnightBlue,urlcolor=MidnightBlue}

\usepackage{svg}

\usepackage{amsmath}

\usepackage{amsmath} 

\theoremstyle{remark}

\newcommand{\prlsection}[1]{\textit{#1.}---}

\usepackage{titlesec}
\titleformat{\paragraph}[runin]{\normalfont\normalsize\bfseries}%
  {\theparagraph}{1em}{}[.]
\titlespacing*{\paragraph}{0pt}{1.5ex}{1em}

\usepackage{amsmath,amssymb}  
\usepackage{mathrsfs} 

\begin{document}

\title{Diffusivity in Dissipative Quantum Transport from an Exactly Solvable Krylov Chain}

\author{Zhi-Li Zhou}
\date{\today}
\email{zhiliz2@illinois.edu}
\affiliation{Illinois Center for Advanced Studies of the Universe\\ Department of Physics, 
University of Illinois at Urbana-Champaign, Urbana, IL 61801, USA}

\author{Jorge Noronha}
\date{\today}
\email{jn0508@illinois.edu}
\affiliation{Illinois Center for Advanced Studies of the Universe\\ Department of Physics, 
University of Illinois at Urbana-Champaign, Urbana, IL 61801, USA}

\begin{abstract}
The computation of transport coefficients in interacting quantum many-body systems is rarely analytically accessible.
Here, we develop a new Krylov-space mechanism that makes the leading density dependent correction to Green-Kubo diffusivity \emph{exactly} calculable in the strong-dissipation regime of a one-dimensional noisy spin-$\frac{1}{2}$ XXZ chain. 
This is done by showing that the density-polarization-dressed bond coherence generates a Krylov subspace where repeated action of the dissipator closes exactly on an explicitly identifiable operator family. Within this subspace, the dissipative dynamics
is then mapped onto a self-similar semi-infinite chain with a boundary defect. Surprisingly, \emph{all} Lanczos coefficients of the
Krylov chain and its boundary Green's function can be determined exactly. The latter then determines the exact leading density-dependent correction to the diffusivity. Finally, we calculate analytically the full boundary-to-bulk Green's function and find that it decays exponentially along the emergent Krylov chain.
\end{abstract}

\maketitle

\prlsection{Introduction} Transport provides a direct link between the microscopic properties and the emergent hydrodynamic behavior displayed by quantum many-body systems. While hydrodynamics constrains the universal infrared behavior of correlation functions \cite{KADANOFF1963419}, it does not determine the associated transport coefficients, which must be derived from the underlying microscopic model \cite{LandauLifshitz1987, Kovtun_2012}. Microscopically, diffusivity is encoded in the zero-frequency current autocorrelation function determined in equilibrium through the Green-Kubo formula \cite{Kubo2012StatisticalPhysicsII}, schematically $D_{\mathrm{GK}}\sim\int_{0}^{\infty} dt\,\langle J(t)J(0)\rangle$. In an interacting system, the determination of $D_{\mathrm{GK}}$ is challenging because the current operator evolves in time as $J(t)=J(0)+it[H,J(0)]+\frac{(it)^2}{2!}[H,[H,J(0)]]+\cdots$, so that successive nested commutators with the Hamiltonian generate an increasingly complicated hierarchy of many-body operators \cite{Rabinovici_2021, PhysRevX.9.041017}, which make evaluating $D_{\mathrm{GK}}$ challenging even using numerical methods \cite{RevModPhys.93.025003}. In fact, even for Bethe-ansatz integrable models, the quantitative evaluation of full finite-temperature current correlation functions and finite-frequency transport responses remains a complex task \cite{RevModPhys.93.025003, De_Nardis_2019, PhysRevLett.122.127202}. Therefore, determining the transport properties in quantum many-body systems, and under what conditions such a problem can become analytically tractable, is arguably one of the main challenges of theoretical physics \cite{RevModPhys.93.025003}.

The spin-$\frac{1}{2}$ XXZ chain provides a canonical platform for addressing this question as this system exhibits rich transport behavior \cite{RevModPhys.93.025003, RevModPhys.94.045006} that can also be investigated experimentally using ultracold atomic systems. In such experiments, the exchange anisotropy can be tuned, and spin transport can be directly probed through the decay of imprinted spin helices \cite{Jepsen_2020}. At the same time, dissipation and stochastic driving have become controllable ingredients in quantum simulation \cite{Barreiro_2011, PhysRevLett.118.140403, muller2012engineeredopensystemsquantum}, and suitably engineered white-noise Hamiltonian couplings give rise to Lindbladian dynamics after averaging over the noise \cite{10.1063/1.1703941, Haken1973, PhysRevLett.118.140403, De_Checchi_2025}. These developments motivate the investigation of hydrodynamic and transport behavior of the XXZ chain subject to controlled stochastic exchange.

In this Letter, we identify a novel mechanism by which the leading interaction-induced correction to Green-Kubo diffusivity becomes \emph{exactly} solvable in the XXZ chain subject to symmetric incoherent nearest-neighbor spin exchange. Using a Jordan-Wigner transformation, this model is exactly mapped onto an interacting spinless-fermion chain \cite{RevModPhys.93.025003} subject to symmetric incoherent nearest-neighbor hopping. The local noise and Hamiltonian preserve the strong $U(1)$ symmetry \cite{Bu_a_2012} and generate diffusive behavior at late times. Additionally, symmetry arguments ensure that $D_{\mathrm{GK}}$ is separately even in the coherent hopping amplitude $J$ and nearest-neighbor interaction strength $U$ (the XXZ anisotropy is $U/2J$), so that the leading interaction-induced correction to diffusion is of order $\mathcal{O}(J^2U^2/\gamma^3)$, when $|J|/\gamma,|U|/\gamma \ll 1$. 

The analytical determination of the leading density dependent contribution to diffusivity, for arbitrary values of $U/J$ in this strong dissipation regime, is the main result of this work. This is accomplished by showing that the repeated action of the dissipator on the density-polarization-dressed bond coherence, $\mathcal{B}_{1}$, closes exactly on an analytically identifiable operator family. Within this subspace, $\mathcal{D}\mathcal{B}_{m}$ obeys a recurrence relation, which gives rise to a self-similar semi-infinite 1D Krylov chain with a single boundary defect. This emergent geometry makes the corresponding Green's function exactly solvable, namely $G_{m1}=(\sqrt{3}-1)(2-\sqrt{3})^{m-1}$. In particular, the boundary Green's function $G_{11}=\sqrt{3}-1$ fully determines the leading density-dependent correction to $D_{\mathrm{GK}}$ coming from interactions in closed form. Furthermore, the exponential decay of $G_{m1}$ with the Krylov coordinate $m$ has an interesting interpretation as the screening of $\mathcal{B}_{m}$ in operator space, as we discuss in this work.

\prlsection{Model and hydrodynamic description} We consider the Lindbladian evolution of a spin-$\frac{1}{2}$ XXZ chain subject to symmetric incoherent nearest-neighbor exchange
\begin{equation}
H_{\mathrm{XXZ}}=2J\sum_{j}\left(S_{j}^{x}S_{j+1}^{x}+S_{j}^{y}S_{j+1}^{y}+\frac{U}{2J}S_{j}^{z} S_{j+1}^{z}\right)
\end{equation}
with jump operators $L_{j,+}=\sqrt{\frac{\gamma}{2}}\,
S_{j}^{+}S_{j+1}^{-}$ and $L_{j,-}=\sqrt{\frac{\gamma}{2}}\,
S_{j+1}^{+}S_{j}^{-}$. Both the coherent and dissipative dynamics locally conserve the total magnetization $S_{\mathrm{tot}}^{z}=\sum_{j}S_{j}^{z}$. Using $S_{j}^{z}=n_{j}-\frac{1}{2}$, we apply a Jordan-Wigner transformation \cite{Coleman_2015}, followed by the staggered gauge transformation $c_{j}\rightarrow(-1)^{j}c_{j}$, to map the model exactly onto the fermionic model at the level of the Lindbladian dynamics, with the jump operators differing only by physically irrelevant overall phases within each fermion-parity sector. Therefore, we obtain $H=H_{\mathrm{hop}}+H_{\mathrm{int}}$ with
\begin{equation}
H_{\mathrm{hop}}=-J\sum_{j}\left(c_{j+1}^\dagger c_{j}+c_{j}^{\dagger}c_{j+1}\right),
\end{equation}
\begin{equation}
H_{\mathrm{int}}=U\sum_{j} \left(n_{j}-\frac{1}{2}\right)\left(n_{j+1}-\frac{1}{2}\right),
\end{equation}
The jump operators in this description are $L_{j,1}=\sqrt{\frac{\gamma}{2}}\,c_{j}^{\dagger}c_{j+1}$ and $L_{j,2}=\sqrt{\frac{\gamma}{2}}\,c_{j+1}^{\dagger} c_{j}$. From now on, we work in the fermionic representation. Note that these transformations are changes of representation and do not alter the transport coefficient. In fact, we remark that $S_{j}^{z}=n_{j}-\frac{1}{2}$ maps the conserved spin density directly onto the fermion density, while the staggered gauge transformation leaves $n_{j}$ invariant. Thus, one can see that the long-wavelength relaxation eigenvalue and the corresponding diffusion coefficient extracted from it remain the same.\footnote{We note that for our periodic chain the Jordan-Wigner transformation gives the parity-dependent boundary condition $c_{L+1}=-p\,c_{1}$ in the sector with fermion parity $p=\pm 1$, which does not affect the diffusion coefficient in the thermodynamic limit.}

Now, we investigate the hydrodynamic behavior of this model. First, the total charge $Q=\sum_{j}n_{j}$ is strongly conserved,
$[Q,H]=[Q,L_{j,\sigma}]=0$, and for $\gamma>0$ the jumps break
the additional local conservation laws of the closed XXZ
chain. Hence, $Q$ is the only independent additive conserved
quantity with a density of fixed finite range (up to the identity). The corresponding density obeys the following exact continuity equation
\begin{equation}
\mathcal{L}^{\dagger}n_{j}=j^{H}_{j-1/2}-j^{H}_{j+1/2}+\frac{\gamma}{2}\left(n_{j-1}-2n_{j}+n_{j+1}\right),
\label{eq:continuity equation}
\end{equation}
where $j^{H}_{j+1/2}=iJ(c_{j+1}^{\dagger}c_{j}-c_{j}^{\dagger}c_{j+1})$ is the coherent current. Equivalently, the stochastic contribution itself can be written as the divergence of $j^D_{j+1/2}=\frac{\gamma}{2}(n_{j}-n_{j+1})$, so that the microscopic current separates into coherent and stochastic pieces, $j_{j+1/2}=j^{H}_{j+1/2}+j^{D}_{j+1/2}$. Fourier transforming Eq.~\eqref{eq:continuity equation} and taking the hydrodynamic limit, we obtain\footnote{We set the lattice spacing to unity and adopt the Fourier convention $n_{k}=\sum_{j}e^{-ikj}n_{j}$, $j_{k}^{H}=\sum_{j}e^{-ik(j+1/2)}j_{j+1/2}^{H}$, with the current centered on the bond $(j,j+1)$.}
\begin{equation}
\mathcal{L}^\dagger(k)n_{k}=-ik\,j_{k}^{H}-\frac{\gamma}{2}k^{2}n_{k}+\mathcal{O}(k^{3}).
\end{equation}
We now argue that the system supports only a diffusive mode. Since $n$ is the only nontrivial local conserved density, the linearized Euler dynamics contains a single velocity, $v_{\mathrm{E}}=\left.\frac{\partial j_{\mathrm{E}}(n)}{\partial n}\right|_{\bar{n}}$, which corresponds to the propagation velocity of the hydrodynamic mode.
However, the system possesses spatial inversion symmetry, which implies $v_{\mathrm{E}}=0$, thus excluding the ballistic hydrodynamic mode. Beyond linear order, nonlinear fluctuating hydrodynamics \cite{Spohn_2014, PhysRevLett.111.230601, PhysRevLett.108.180601} is controlled by the quadratic Euler coupling $G=\frac{1}{2}\left.\frac{\partial^{2}j_{\mathrm{E}}(n)}{\partial n^{2}}\right|_{\bar{n}}$. In the present model, the current appearing in the exact continuity equation contains no nonlinear density coupling such as $n_jn_{j+1}$, and hence $G=0$ (consistent with spatial inversion symmetry). Finally, since the remaining current sector relaxes at a finite rate, locality implies that diffusion dominates the long-wavelength dynamics, $\lambda(k)=-Dk^{2}+\mathcal{O}(k^{4})$, where $\lambda(k)$ is the Lindbladian eigenvalue continuously connected to the conserved density mode at $k=0$.

We note that the continuity equation exhibits two physically distinct contributions to $D$. The stochastic exchange directly produces the current $j^{D}=-\frac{\gamma}{2}\nabla n$, and therefore the bare contribution $D_{\mathrm{stoch}}=\frac{\gamma}{2}$. The coherent current, by contrast, is not conserved. Its excursions through the relaxing sector generate an additional zero-frequency transport contribution. Hence, we write $D=\frac{\gamma}{2}+D_{\mathrm{GK}}$, where the two distinct contributions are additive \cite{hippert2021universalmanybodydiffusionmomentum}. The first term is fixed locally by the jump process, whereas all nontrivial dependence on $J$ and $U$ is contained in $D_{\mathrm{GK}}$.

We now formulate this contribution in terms of current correlations
in the system's stationary state. First, the Jaynes maximum entropy principle \cite{Jaynes:1957zza} selects the grand-canonical family $\rho_{\mathrm{eq}}=e^{\mu Q}/Z$ at fixed mean charge. Here $\mu$ is the Lagrange multiplier conjugate to $Q$, with $\bar{n}=\operatorname{Tr}(\rho_{\mathrm{eq}}n_{j})=\frac{e^\mu}{1+e^{\mu}}$. It is easy to verify that the maximum entropy state is also an exact stationary state of Lindbladian dynamics, namely $\mathcal{L}(\rho_{\mathrm{eq}})=0$. The natural equilibrium correlation for linear response is the Bogoliubov-Kubo-Mori (BKM) inner product \cite{Mori:1965oqj, Kubo1966}, which is defined as $\langle A,B\rangle_{\mathrm{BKM}}=\int_{0}^{1} ds\,\operatorname{Tr}\left(\rho_{\mathrm{eq}}^{1-s} A^{\dagger}\rho_{\mathrm{eq}}^{s}B\right)$. We define the charge fluctuation as $\Delta Q=Q-\langle Q\rangle_{\mathrm{eq}}$, the corresponding (extensive) BKM static susceptibility is given by $\chi=\langle\Delta Q,\Delta Q\rangle_{\mathrm{BKM}}$. Since $\Delta Q$ commutes with $\rho_{\mathrm{eq}}$, this reduces to the equilibrium charge variance. Furthermore, because $\rho_{\mathrm{eq}}$ is a product Bernoulli state, we find $\chi=\langle(\Delta Q)^2\rangle_{\mathrm{eq}}=L\bar{n}(1-\bar{n})$. 
Finally, we define the total coherent current $J^H=\sum_{j} j^{H}_{j+1/2}$ and below we express its contribution to diffusion using the projection-operator form of the Green-Kubo formula.

Following the Kubo-Mori construction \cite{Mori:1965oqj, Kubo2012StatisticalPhysicsII}, let
\begin{equation}
\mathcal{P}A=\frac{\langle\Delta Q,A\rangle_{\mathrm{BKM}}}{\chi}\,\Delta Q, 
\quad 
\mathcal{Q}=\mathbf{1}-\mathcal{P},
\end{equation}
where $\mathcal{P}$ is the projector onto the hydrodynamic subspace generated by the conserved charge fluctuation, and $\mathcal{Q}$ projects onto its complement. Applying the Kubo-Mori projected dynamics to Lindbladian evolution, we obtain (see End Matter for details) the dynamical contribution to the diffusion coefficient as
\begin{equation}
D_{\mathrm{GK}}=\frac{1}{\chi}\int_{0}^{\infty}dt\, \left\langle \mathcal{Q}J^{H},e^{t\mathcal{Q}\mathcal{L}^\dagger\mathcal{Q}}\mathcal{Q}J^{H}\right\rangle_{\mathrm{BKM}}.
\label{eq:GK general}
\end{equation}
Note that only the component of the current orthogonal to the hydrodynamic subspace is propagated, and it evolves entirely within the relaxing sector. 
For the present model, one can employ spatial inversion to simplify Eq.~\eqref{eq:GK general}. Since $\Delta Q$ is even under inversion whereas $J^H$ is odd,
we obtain $\langle\Delta Q,J^H\rangle_{\mathrm{BKM}}=0$. Hence, $\mathcal{P}J^{H}=0$ and $\mathcal{Q}J^{H}=J^{H}$. One can then show that the time integral can equivalently be written as
\begin{equation}
    D_{\mathrm{GK}}=\frac{1}{\chi}\left\langle J^{H},-\left(\mathcal{Q}\mathcal{L}^{\dagger}\mathcal{Q}\right)^{-1}J^{H}\right\rangle_{\mathrm{BKM}}.
    \label{eq:GK resolvent form}
\end{equation}
When $U=0$, the calculation of Eq.~\eqref{eq:GK general} simplifies \cite{Eisler_2011, Penc_2026} since $\mathcal{L}^{\dagger}J^{H}=-\gamma J^{H}$, so $D_{\mathrm{GK}}=2J^{2}/\gamma$. For $U\neq0$, however, the interaction rotates $J^{H}$ into an infinite hierarchy of operators. Nevertheless, one can prove quite generally that 
\begin{equation}
    0\leq D_{\mathrm{GK}}\leq\frac{2J^{2}}{\gamma},
\end{equation}
so the inequality is saturated by the non-interacting result. Thus, excursions induced by interactions
can only reduce $D_{\mathrm{GK}}$ relative to its free value.
The central question now is whether the leading interaction
correction in the strong dissipation regime can be computed
analytically. We show that this is indeed possible below.

\prlsection{Symmetry-constrained expansion} Before solving Eq.~\eqref{eq:GK resolvent form}, we note that its dependence on $J$ is strongly constrained by exact symmetries. First, the staggered gauge transformation $Sc_{j}S^{-1}=(-1)^{j}c_{j}$ leaves the dissipator invariant while mapping $H(J,U)\to H(-J,U)$ and $J^{H} \to -J^{H}$. Hence, one obtains $D_{\mathrm{GK}}(J,U)=D_{\mathrm{GK}}(-J,U)$. Second, consider complex conjugation $K$ in the occupation basis. Since $KHK^{-1}=H$, $KiK^{-1}=-i$, $K$ reverses $i[H,\cdot]$. Now, we define the antiunitary transformation $\Theta=SK$, which satisfies $\Theta \mathcal{L}^{\dagger}(J,U)\Theta^{-1}=\mathcal{L}^{\dagger}(J,-U)$ and $\Theta J^{H}\Theta^{-1}=J^{H}$. Together with the invariance of $\rho_{\mathrm{eq}}$, and the fact that antiunitarity
complex-conjugates the BKM current correlation but the total current and its projected evolution
are Hermitian, we obtain $D_{\mathrm{GK}}(J,U)=D_{\mathrm{GK}}(J,-U)$. Therefore, $D_{\mathrm{GK}}$ is separately even in $J$ and $U$, as physically expected in the spinless-fermion representation.

Since $J^{H}$ is odd under lattice reflection while conserved charge is even, and the full Lindbladian preserves reflection parity, the entire current-generated sector remains orthogonal to the hydrodynamic subspace. Hence, $\mathcal{Q}=\mathbf{1}$ on this sector and $\mathcal{Q}\mathcal{L}^{\dagger}\mathcal{Q}=\mathcal{L}^{\dagger}$. Therefore, we may write $-\mathcal{L}^{\dagger}=\gamma \mathcal{D}+\mathcal{L}_{H}$, where $\mathcal{L}_{H}=-i[H,\cdot]$ and $\mathcal{D}$ is the dimensionless dissipator. In the strong-dissipation regime, $\gamma \gg |J|,|U|$, Eq.~\eqref{eq:GK resolvent form} admits the following Neumann series expansion
\begin{equation}
\begin{aligned}
(\gamma\mathcal{D}+\mathcal{L}_H)^{-1}
&=\frac{1}{\gamma}\mathcal{D}^{-1}-\frac{1}{\gamma^{2}}\mathcal{D}^{-1}\mathcal{L}_H\mathcal{D}^{-1}\\
&\quad
+\frac{1}{\gamma^{3}}\mathcal{D}^{-1}\mathcal{L}_H\mathcal{D}^{-1}
\mathcal{L}_{H}\mathcal{D}^{-1}+\cdots.
\end{aligned}
\end{equation}
Correspondingly, we have
$D_{\mathrm{GK}}=D_{\mathrm{GK}}^{(1)}
+D_{\mathrm{GK}}^{(2)}+D_{\mathrm{GK}}^{(3)}+\cdots$.
Using $\mathcal{D}J^H=J^H$ and
$\langle J^{H},J^{H}\rangle_{\mathrm{BKM}}=2\chi J^{2}$,
we obtain $D_{\mathrm{GK}}^{(1)}=2J^{2}/\gamma$.
We note that all even inverse powers of $\gamma$ vanish as their alternating operator products are BKM skew-adjoint and preserve Hermiticity. In particular, $D_{\mathrm{GK}}^{(2)}=0$, consistent with the exact $U\to -U$ symmetry. The leading nonvanishing interaction-induced correction therefore appears at third order, which is given by
\begin{equation}
D_{\mathrm{GK}}^{(3)}=-\frac{1}{\chi\gamma^{3}}\left\langle \mathcal{L}_{H}J^{H},\mathcal{D}^{-1}\mathcal{L}_{H}J^{H}\right\rangle_{\mathrm{BKM}}.
\end{equation}
Since $[H_{\mathrm{hop}},J^{H}]=0$, only the interaction dresses the current. Defining the bond coherence operator $h_{j}=c^{\dagger}_{j+1}c_{j}+c^{\dagger}_{j}c_{j+1}$, one finds $\mathcal{L}_{H}J^{H}=JU\mathcal{B}_{1}$, $\mathcal{B}_{1}=\sum_{j}(n_{j+2}-n_{j-1})h_{j}$. Therefore, the expansion reduces to
\begin{equation}
    D_{\mathrm{GK}}=\frac{2J^2}{\gamma}-\frac{J^2U^2}{\chi \gamma^3}\langle \mathcal{B}_{1},\mathcal{D}^{-1}\mathcal{B}_{1}\rangle_{\mathrm{BKM}}+\mathcal{O}({\gamma^{-5}}).
\end{equation}
The entire many-body problem at this order reduced to a single Green's function $\langle \mathcal{B}_{1},\mathcal{D}^{-1}\mathcal{B}_{1}\rangle_{\mathrm{BKM}}$. 
As we now show, the Krylov construction generated from $\mathcal{B}_{1}$ closes exactly on a simple operator family under $\mathcal{D}$. This reduces the many-body inverse problem to the boundary Green's function of a semi-infinite 1D Krylov chain.

\prlsection{Emergent Krylov chain} We now define $\mathcal{B}_{m}=\sum_{j}(n_{j+m+1}-n_{j-m})h_{j}$, where $m$ measures the separation of the density dressing from the bond. Remarkably, the action of $\mathcal{D}$ on $\mathcal{B}_{m}$ closes exactly \footnote{We note that, at finite even $L$, the operators with $1\leq m\leq L-2$ satisfy $\mathcal{B}_{L-1-m}=-\mathcal{B}_{m}$.}
\begin{equation}
    \mathcal{D}\mathcal{B}_{1}=\frac{3}{2}\mathcal{B}_{1}-\frac{1}{2}\mathcal{B}_{2},
\end{equation}
\begin{equation}
    \mathcal{D}\mathcal{B}_{m}=2\mathcal{B}_{m}-\frac{1}{2}\mathcal{B}_{m-1}-\frac{1}{2}\mathcal{B}_{m+1},\quad m\geq2.
\end{equation}
The special boundary coefficient $\frac{3}{2}$ originates from Lindblad jumps that overlap both the density dressing and the bond coherence. Away from the boundary, this overlap is absent and the recurrence becomes translationally invariant. These relations imply $\mathcal{K}(\mathcal{B}_{1};\mathcal{D})=\overline{\mathrm{span}}\{\mathcal{B}_{1},\mathcal{D}\mathcal{B}_{1},\mathcal{D}^{2}\mathcal{B}_{1},\cdots\}=\overline{\mathrm{span}}\{\mathcal{B}_{m}\}_{m\geq1}$. In the thermodynamic limit, these operators are mutually orthogonal in the BKM inner product, i.e. $\lim_{L\to\infty}\frac{1}{L}
\langle\mathcal{B}_{m},\mathcal{B}_{m'}\rangle_{\mathrm{BKM}}
=4[\bar{n}(1-\bar{n})]^2\delta_{mm'}$. Hence, we define the normalized Krylov basis $|m)=\frac{\mathcal{B}_{m}}{2\sqrt{L}\,\bar{n}(1-\bar{n})}$, with inner products given by the thermodynamic limits of
the finite-chain BKM inner products, so that $(m'|m)=\delta_{m'm}$. Then, the completed Krylov subspace can be identified unitarily with a single-particle Hilbert space, namely $\mathcal{U}:\mathcal{K}(\mathcal{B}_{1};\mathcal{D})\to \ell^{2}(\mathbb{N})$ and $\mathsf{D}=\mathcal{U}\mathcal{D}\,\mathcal{U}^{-1}$ is the matrix representation of $\mathcal{D}$ on the Krylov chain. Therefore, the above recurrence relation becomes
\begin{equation}
    \mathsf{D}|1)=\frac{3}{2}|1)-\frac{1}{2}|2),
\end{equation}
\begin{equation}
    \mathsf{D}|m)=2|m)-\frac{1}{2}|m-1)-\frac{1}{2}|m+1),\quad m\geq 2.
\end{equation}
The system then becomes a semi-infinite 1D Krylov chain with a single boundary defect. We remark that the Krylov representation of operator dynamics as one 1D chains are standard in recursion method \cite{viswanath_2025_avfvz-v0s16}. However, for generic interacting many-body systems, only a finite number of Lanczos coefficients can typically be computed explicitly, and the remaining semi-infinite Krylov tail must be modeled using asymptotic information \cite{PhysRevX.9.041017, Pinna_2025, PhysRevB.110.104413}. In contrast, for our system, $\{|m)\}$ are precisely the Lanczos basis generated from $|1)$. The corresponding Lanczos coefficients are therefore known exactly for the entire semi-infinite chain, $a_{1}=\frac{3}{2}$, $a_{m}=2\,(m\geq2)$, $b_{m}=\frac{1}{2}\,(m\geq1)$, so no truncation of the Krylov tail is required.

Now we proceed to calculate the boundary Green's function of the Krylov chain. We define $G_{mn}=(m|\mathsf{D}^{-1}|n)$, where $\sum_{\ell \geq1}\mathsf{D}_{m\ell}G_{\ell n}=\delta_{mn}$. Thus, we obtain $\langle \mathcal{B}_{1},\mathcal{D}^{-1}\mathcal{B}_{1}\rangle_{\mathrm{BKM}}=4L[\bar{n}(1-\bar{n})]^{2}G_{11}$, where $G_{11}=(1|\mathsf{D}^{-1}|1)=\int^{\infty}_{0}dt\,(1|e^{-t\mathsf{D}}|1)$. Physically, $G_{11}$ is the time-integrated return propagator at the boundary site and exactly resums excursions through the entire infinite hierarchy. To evaluate $G_{11}$, we separate the boundary site from the homogeneous bulk $\ell^{2}(\mathbb{N})=\mathrm{span}\{|1)\}\oplus
\overline{\mathrm{span}\{|2),|3),\cdots\}}$,
and we obtain
\begin{equation}
\mathsf{D}=\begin{pmatrix} a_{1} & -b(2|\\ -b|2) & \mathsf{D}_{\mathrm{bulk}} \end{pmatrix}, \quad a_1=\frac{3}{2},\, b=\frac{1}{2} .
\end{equation}
Then, the Schur complement gives
\begin{equation}
    G_{11}=\frac{1}{a_{1}-b^{2}T},\quad T=(2|\mathsf{D}^{-1}_{\mathrm{bulk}}|2).
\end{equation}
Since the bulk has diagonal $a=2$ and is \emph{self-similar},
integrating out one more site yields the fixed-point equation
\begin{equation}
    T=\frac{1}{a-b^{2}T}=\cfrac{1}{a-\cfrac{b^{2}}{a-\cfrac{b^{2}}{a-\dots}}}.
\end{equation}
The continued-fraction form above shows that each level of the recursion corresponds to integrating out one additional site, while its self-similar structure exactly resums propagation through the entire infinite Krylov chain and equivalently encodes it into a boundary self-energy $b^{2}T$. The physical solution compatible with the positive bulk spectrum (which gives $0<T\leq1$) is $T=4-2\sqrt{3}$ and, therefore, $G_{11}=\sqrt{3}-1$.

In addition, we can compute $G_{m1}$ from the same recurrence, which is given by $G_{m1}=(\sqrt{3}-1)(2-\sqrt{3})^{m-1}$, $m\geq 1$. Thus, we remark that the response generated at the boundary is exponentially localized along the Krylov coordinate, namely $G_{m1}\sim e^{-(m-1)/\xi_{\mathrm{K}}}$, where the inverse screening length is given by $\xi_{\mathrm{K}}^{-1}=\ln(2+\sqrt{3})$. One may refer to this mechanism as \emph{Krylov-space polarization screening} because a polarization source, $\mathcal{B}_{1}$ at the boundary $m=1$, generates a zero-frequency response $G_{m1}$ that decays exponentially with distance along the Krylov chain, thus defining a finite screening length, $\xi_{\mathrm{K}}\approx 0.759$. 

Therefore, putting these results together, we find
\begin{equation}
 D_{\mathrm{GK}}=\frac{2J^{2}}{\gamma}-\frac{4(\sqrt{3}-1)J^{2}U^{2}\bar{n}(1-\bar{n})}{\gamma^{3}}+\mathcal{O}(\gamma^{-5}),
 \label{eq:finalresult}
\end{equation}
in the thermodynamic limit of the strong-dissipation regime. More broadly, Eq.~\eqref{eq:finalresult} constitutes a rare exact microscopic determination of the density dependence of a transport coefficient in an interacting open quantum many-body system and its solvability originates from the exact resummation of an infinite operator hierarchy through the emergent Krylov chain. Finally, we remark that our result also provides a closed form microscopic diffusivity for macroscopic fluctuation theory (MFT) \cite{RevModPhys.87.593} and, as such, it therefore provides a rigorous benchmark for recent efforts to derive nonlinear fluctuating hydrodynamics and MFT directly from interacting noisy quantum matter \cite{Bernard:2026dnc, Christopoulos:2026jze}.

\prlsection{Conclusion} In this work, we showed for the first time how an infinite hierarchy of density-dressed coherences becomes an exactly solvable emergent Krylov chain whose boundary response determines the leading density dependent
correction to Green-Kubo diffusivity in the strong-dissipation regime of a spin-$\frac{1}{2}$ XXZ chain with symmetric incoherent spin exchange. Our approach is distinct from previously established approaches to exact solvability in Lindbladian dynamics \cite{Fazio_2025, Penc_2026, PhysRevLett.107.137201, Prosen_2008, PhysRevLett.122.010401, PhysRevE.102.062210, PhysRevB.109.L140302, Barthel_2022, PhysRevLett.117.137202, Ziolkowska_2020, PhysRevLett.120.090401, PhysRevLett.119.190402} in that it is rooted in the exact closure of the dissipative dynamics onto a Krylov chain with a simple self-similar geometry. This, in principle, suggests a novel route to obtaining exact results for transport in interacting open quantum systems, without requiring the full many-body Lindbladian itself to be exactly solvable.

Furthermore, we note that our method resolves the infinite dissipative operator hierarchy associated with the leading interaction correction by summing all excursions within this hierarchy when $|J|/\gamma,|U|/\gamma\ll1$, for arbitrary values of $U/J$. Therefore, at half filling, our result covers parameter regimes of the closed XXZ chain corresponding to both gapless and gapped ground states.

Interestingly, our calculations suggest the presence of a Krylov-space polarization screening mechanism characterized by a finite screening length in operator space in which density-dressed coherence is screened while the system's total charge remains conserved. Physically, this finite screening length along the emergent Krylov chain stems from the competition between the spreading of the interaction-induced density dressing  and the damping of the hopping coherence, which are both caused by our symmetric noise. We remark that this is different from Krylov space localization \cite{Rabinovici:2021qqt, PhysRevResearch.5.033085} since our dissipation-induced screening attenuates the response without localizing
the underlying Krylov modes.

Finally, our work motivates a broader strategy in which one could use the structure of the noise itself to unveil a hidden solvable hierarchy in Krylov space involving the transport of other interacting open quantum systems. In this regard, it would be interesting to investigate other systems that may share some of the features of the model considered here, such as fermion chains with finite-range density interactions and, in particular, Fermi-Hubbard chains with symmetric incoherent hopping of each spin species. We hope to report on such systems in the near future. 

\begin{acknowledgments}
\prlsection{Acknowledgments}The authors are partly supported by the U.S. Department of Energy, Office of Science, Office of Nuclear Physics under Award No. DE-SC0023861.
\end{acknowledgments}

\bibliography{refs}

\appendix

\section{End Matter}

\prlsection{Derivation of diffusion coefficient} Here we derive Eq.~\eqref{eq:GK general} from the relaxation of a long-wavelength density perturbation. Following \cite{Kubo2012StatisticalPhysicsII}, we consider a weak density modulation of density operator
\begin{equation}
    \rho_{h}=\frac{\exp{(\mu Q+h_{k}n_{-k}+h_{-k}n_{k})}}{Z[h]}.
\end{equation}
To first order in $h_{k}$, we obtain $\delta\langle n_{k}(0)\rangle=\chi_{k}h_{k}$. After the perturbation is removed and the system evolves with the unperturbed Lindbladian dynamics, one finds $\delta\langle n_{k}(t)\rangle=h_{k}\chi_{k}C_{nn}(k,t)$, where $C_{nn}(k,t)$ is the normalized density-density correlation function 
\begin{equation}
    C_{nn}(k,t)=\frac{1}{\chi_{k}}\langle n_{k}, e^{t\mathcal{L}^{\dagger}(k)}n_{k}\rangle_{\mathrm{BKM}}.
\end{equation}
Hence, the slowest pole of $C_{nn}$ controls the decay of an infinitesimal density perturbation. Fourier transforming the continuity equation Eq.~\eqref{eq:continuity equation} to better study its long-wavelength behavior, we find the following exact equation
\begin{equation}
    \mathcal{L}^{\dagger}(k)n_{k}=-i2\sin\left(\frac{k}{2}\right)j^{H}_{k}-2\gamma\sin^{2}\left(\frac{k}{2}\right)n_{k}.
    \label{eq: k-continuity equation}
\end{equation}
Note that the stochastic current, $j^{D}_{k}=-i\gamma\sin(\frac{k}{2})n_{k}$, which is generated by jump operators, lies entirely within the hydrodynamic subspace. 
According to the Mori-Zwanzig formalism \cite{Mori:1965oqj, 10.1093/oso/9780195140187.001.0001}, we define the following finite-$k$ hydrodynamic projector
\begin{equation}
    \mathcal{P}_{k}(A)=n_{k}\frac{\langle n_{k},A\rangle_{\mathrm{BKM}}}{\chi_{k}}, \quad \mathcal{Q}_{k}=\mathbf{1}-\mathcal{P}_{k}.
\end{equation}
Therefore, we can derive the exact Mori equation for the density-density correlator, and using Eq.~\eqref{eq: k-continuity equation}, we obtain
\begin{equation}
\begin{aligned}
&\partial_{t}C_{nn}(k,t)=-\frac{\gamma}{2}
 \left(2\sin\frac{k}{2}\right)^{2}C_{nn}(k,t)\\
&\qquad
-\left(2\sin\frac{k}{2}\right)^{2}
\int_{0}^{t}ds\,K(k,s)C_{nn}(k,t-s),
\label{eq: Mori equation}
\end{aligned}
\end{equation}
where the memory kernel is defined as
\begin{equation}
    K(k,t)=\frac{1}{\chi_{k}}\langle \mathcal{Q}_{k}j^{H}_{k},e^{t\mathcal{Q}_{k}\mathcal{L}^{\dagger}\mathcal{Q}_{k}}\mathcal{Q}_{k}j^{H}_{k}\rangle_{\mathrm{BKM}}.
\end{equation}
We remark that, on the right hand side of Eq.~\eqref{eq: Mori equation}, the first term represents the diffusion created by jump operators, the second term is the coherent-current memory contribution determining Green-Kubo diffusion, and there are no ballistic modes.

By performing a Laplace transform of the Mori equation, we obtain
\begin{equation}
\widetilde{C}_{nn}(k,z)=\frac{1}{z+(2\sin \frac{k}{2})^{2}\left [\frac{\gamma}{2}+\widetilde{K}(k,z)\right]}.
\end{equation}
Hence, one can see that the poles are determined by the denominator of the above equation.

Now, we are ready to discuss the hydrodynamic behavior of the theory (we note that we take the thermodynamic limit before the hydrodynamic limit). Because the fast modes are stable (gapped in the linear hydrodynamic regime), we expect that  $\int_{0}^{\infty}dt\,|K(k,t)|<\infty$ uniformly for sufficiently small $k$. On the hydrodynamic time scale $t\sim k^{-2}$, the memory kernel then decays rapidly compared with $C_{nn}(k,t)$, so
\begin{equation}
    \int_{0}^{t}ds\,K(k,s)C_{nn}(k,t-s)\simeq C_{nn}(k,t)\int_{0}^{\infty}ds\,K(0,s).
\end{equation}
Consequently, the fast sector remains regular as $k\to0$ when $z\to0^{+}$, the corresponding iterated limit exists, and is given by $\widetilde{K}(0,0)$. Thus, the exact Mori equation has a diffusive limit, with diffusion coefficient given by
\begin{equation}
    D=\frac{\gamma}{2}+\lim_{z\to 0^{+}}\lim_{k\to 0}\widetilde{K}(k,z):=\frac{\gamma}{2}+D_{\mathrm{GK}},
\end{equation}
where $D_{\mathrm{GK}}$ is determined by the projected Green-Kubo formula
\begin{equation}
\begin{aligned}
    D_{\mathrm{GK}}&=\lim_{z\to 0^{+}}\lim_{k\to 0}\frac{1}{\chi_{k}}\langle \mathcal{Q}_{k}j^{H}_{k},\frac{1}{z-\mathcal{Q}_{k}\mathcal{L}^{\dagger}\mathcal{Q}_{k}}\mathcal{Q}_{k}j^{H}_{k}\rangle_{\mathrm{BKM}}\\
    &=\frac{1}{\chi}\int_{0}^{\infty}dt\, \left\langle \mathcal{Q}J^{H},e^{t\mathcal{Q}\mathcal{L}^\dagger\mathcal{Q}}\mathcal{Q}J^{H}\right\rangle_{\mathrm{BKM}}.
\end{aligned}
\end{equation}

\prlsection{Exact bounds on Green-Kubo diffusivity}
For non-negative $z$, let us define the quantity
\begin{equation}
D_{\mathrm{GK},L}(z)=\frac{1}{\chi}\int_{0}^{\infty}dt\,e^{-zt}
\langle J^{H},e^{t\mathcal{L}^{\dagger}}J^{H}\rangle_{\mathrm{BKM}}.
\end{equation}
Thus, our Green-Kubo diffusion coefficient can be obtained via $D_{\mathrm{GK}}
=\lim_{z\to0^{+}}\lim_{L\to\infty}D_{\mathrm{GK},L}(z)$. 
It is convenient to define an operator $X_{z}$ via
\begin{equation}
(z+\gamma\mathcal{D}+\mathcal{L}_{H})X_{z}=J^{H},
\end{equation}
with which one can see that $D_{\mathrm{GK},L}(z)=\frac{1}{\chi}\langle J^{H},X_{z}\rangle_{\mathrm{BKM}}$. Now, because
\begin{equation}
\mathcal{D}J^{H}=J^{H},\qquad
\langle J^{H},J^{H}\rangle_{\mathrm{BKM}}=2\chi J^{2} 
\end{equation}
for hopping amplitude $J\neq0$, one can always write
\begin{equation}
X_{z}=\tau_{z}J^{H}+Y_{z},\qquad
\langle J^{H},Y_{z}\rangle_{\mathrm{BKM}}=0,
\end{equation}
with real $\tau_{z}$ so that
$D_{\mathrm{GK},L}(z)=2J^{2}\tau_{z}$. Hence, 
taking the BKM inner product of the response
equation with $X_{z}$ then gives
\begin{equation}
\tau_{z}=(z+\gamma)\tau_z^{2}+
\frac{\langle Y_{z},(z+\gamma\mathcal D)Y_{z}
\rangle_{\mathrm{BKM}}}{2\chi J^{2}}.
\end{equation}
The last term is nonnegative because $\mathcal{D}$ is positive semidefinite using the BKM inner product, which implies that 
$0\leq\tau_z\leq(z+\gamma)^{-1}$ and, thus,
\begin{equation}
0\leq D_{\mathrm{GK},L}(z)
\leq\frac{2J^{2}}{z+\gamma}.
\end{equation}
Therefore, we see that
\begin{equation}
0\leq D_{\mathrm{GK}}\leq\frac{2J^{2}}{\gamma},
\end{equation}
where the upper bound is saturated when $U=0$.

\end{document}